\documentclass[twocolumn]{aastex62}
\graphicspath{{./}{figures/}}
\received{...}
\revised{...}
\accepted{...}

\usepackage{natbib}
\usepackage{graphicx}
\usepackage{dcolumn}
\usepackage{bm}

\usepackage{wrapfig}
\usepackage{hyperref}
\usepackage{xcolor}
\hypersetup{
    colorlinks,
    linkcolor={red!50!black},
    citecolor={blue!50!black},
    urlcolor={blue!80!black}
}

\shorttitle{Microlensing path parametrization}
\shortauthors{de Almeida and do Nascimento Jr.}

\begin{document}

\title{Microlensing path parametrization  for Earth-like Exoplanet detection around solar mass stars}


\author{L. de Almeida}\email{dealmeida.l@fisica.ufrn.br}
\affiliation{Dep. de F\'isica, Univ. Federal do Rio Grande do Norte\\  
59072-970 Natal, RN, Brazil}

\author{J.-D.~do Nascimento Jr.}
\affiliation{Dep. de F\'isica, Univ. Federal do Rio Grande do Norte\\  
59072-970 Natal, RN, Brazil} 

\affiliation{Harvard-Smithsonian Center for Astrophysics\\
60 Garden St., Cambridge, MA 02138, USA}

\affiliation{Université Paris-Sud, CNRS, Institut d'Astrophysique\\ 
Spatiale,  UMR 8617, 91405, Orsay Cedex, France}

\begin{abstract}
We propose a new  parametrization of the impact parameter $u_0$ and impact angle $\alpha$ for microlensing systems composed by an Earth-like Exoplanet  around a 
Solar mass Star at 1~AU. We present  the caustic topology of such system,
as well as the related light  curves generated by using such a new parametrization.
Based on   the same density of points and accuracy of regular methods, we obtain
results  5 times faster for discovering Earth-like exoplanet.  In this big data revolution of photometric astronomy, our method will impact future missions like  WFIRST (NASA)  and 
Euclid (ESA) and they data pipelines, providing a rapid and deep detection of exoplanets for this specific  class of microlensing event that might  otherwise be lost. 
\end{abstract}

\keywords{microlensing technique, exoplanets, close topology, path parametrization}

\section{Introduction}\label{sec:intro}

Gravitational lensing of a point source creates two images with combined 
brightness exceeding that of the source. For small separation  between the two 
images the only observable  consequence of the lensing is an apparent  source brightness  variation. This phenomenon is referred as gravitational microlensing
\citep{1936Sci....84..506E,1964PhRv..133..835L,1986ApJ...304....1P,1991ApJ...374L..37M}. 
Gravitational microlensing, among other things, is used as a constraint for 
several questions in astrophysics and  cosmology, as for example to study 
primordial black holes \citep{2011PhRvL.107w1101G} and galaxy dark matter
halo \citet{1995PhRvL..74.2867A}. Simultaneously, the study of exoplanets has grown since the discovery of the first exoplanet orbiting a sun-like star \citep{1995Natur.378..355M} and among several branches the study of habitability 
\citep{2011IAUS..276..349B,2016ApJ...820L..15D} has become one of the most active stellar astrophysics subjects.  Currently, a new surprisingly successful application  concerning microlensing is its capability to finding  furthest and smallest planets  outside the snow line region as compared to any  available 
extrasolar planets detection method \citep{1992ApJ...396..104G,1996ApJ...472..660B}. 
The gravitational microlensing detections made so far present a  variety of 
binary systems, and the detection sensitivity for semimajor axis ranges from $0.5$ AU to $10$ AU and the medium mass of the host star is $0.35 M_\odot$ \citep{2012Natur.481..167C}. 
For these systems, the mass ratio, q, between  the planet ($m_2$) and host star ($m_1$), 
$q = m_2 / m_1$  is higher  than  $1 \times {10}^{-4}$. to date eight microlensing planets with planet-hots 
mass ratio   $q < 1 \times {10}^{-4}$  have been characterized \citep{2018AcA....68....1U}. Gravitational microlensing  is directly sensitive to the ratio of the masses of the planets and its host star, and the 
light curve  give us the projected apparent \textbf{semimajor} axis for the system normalized to the Einstein radius. 

\vspace{0.3cm}

From the observational side, the surveys Microlensing Planet Search (MPS) \citep{1999astro.ph..9433R} and  Microlensing Observations in Astrophysics 
(MOA) \citep{rhi00,2003ApJ...591..204S} demonstrated for the first time that microlensing technique is sensitive enough to detect 
earth-mass exoplanets.

\citet{2017ApJ...840L...3S}  show the possibility to detect  Earth-mass Planet in 
a 1 AU Orbit around an Ultracool Dwarf and  \citet{2009ApJ...703.2082Y} \textbf{present an}
extreme magnification microlensing event and its sensitivity to planets with masses as small as $0.2 M_{\oplus} \simeq 2M_{Mars}$ with projected separations near the Einstein ring (~3 AU).   \citet{2014Sci...345...46G} even showed the capability of microlensing technique to discover Earth-mass planets around 1~AU in binary systems.  
As discussed by \citet{alb01,gau02}, more than $77\%$ of  exoplanetary systems discovered with microlensing techniques  shows planets with masses lower than Jupiter mass and 
with \textbf{semimajor} axis between 1.5 and 4~AU.  These results are consistent with the fact that 
massive planets far away from their central stars are easier to be detected with   microlensing method \citep{sum06,han06}.   
In this context, \cite{1986ApJ...304....1P}  shows that  detection is function of  the
 impact parameter $u_0$ and the impact angle $\alpha$. Here, in  this 
 study we propose a parametrization of the source's path to force it to cross the 
 Caustic Region Of INterest (CROIN by \citealt{penny14}).  This offers an advantage for detecting Earth-like planets around Solar-like stars during microlensing events. 
 
 In Section \ref{section2} we describe the lens equation and the semi-analytic method.
We explore  the caustic topology  for events with a \textbf{semimajor} axis of about 1~AU, with the lens at 7.86 kpc and source at 8 kpc in Section \ref{section3} and explore the close systems topology geometry in Section \ref{section3.1} as well describe our parametrization proposal. We  present light curves where it is possible to conduct an analysis of  the $u_0$ and $\alpha$  variation as a function of a fixed parameters in the lens-planet apparent separation in Section \ref{section3.2}. We constructed a model  to simulate our system based on a semi-analytical method for solving the binary lens equation to take into account the source, lenses, caustic, critic curves and producing images and light curves. We present the resume of our simulations and discussion of our results in Section \ref{summary}.

\section{The Lens equation}
\label{section2}

A gravitational microlensing event occurs when a star in the foreground (lens) passes near the line of sight of a background star (source) and \textbf{thereby bends the source light} from the
original path. This bending of the light generates a relative magnification of the source and if the system source-lens have relative movements, a characteristic light curve is produced.  The deflection of the light by a single star can be \textbf{expressed} by $\alpha = \frac{4GM}{c^{2} r}$, where $\alpha$ is the deflection angle, $M$ é the lens mass, $G$ is the universal gravitational constant, $c$ is the speed of light and $r$ is the 
{\bf impact parameter}. If we establish $D_S$ as the distance between the observer and the source and $D_L$ as the distance between the observer and the lens, we can write  the distance between the source and lens as  $(D_S - D_L)$, and we can  derive the \textbf{well known} equation of the Einstein Radius

\begin{equation}
\theta_E = \sqrt{\frac{4GM}{c^2}\frac{D_S - D_L}{D_L  D_S}}.
\label{einsteinradius}
\end{equation}

The equation \ref{einsteinradius} holds regardless on the alignment between the source and the lens, but if they are aligned, we have the so called Einstein ring. Introducing the 
small distance $\beta$ between the source and the lens, we can derive the lens equation for the single lens case as
$\beta = \theta - \frac{\theta_{E}^{2}}{\theta}$ which is the well known lens equation for the single lens case, and it can be easily solved as a second degree polynomial.

\subsection{Formalism} 

 For the binary-lens case, we can rewrite $\beta$, originally written for single lens case,  using the complex notation to denote the lens equation for the  two lenses \citep{wit90,wit95} case, representing a host star and their planet as

\begin{equation}
\omega = z - \frac{\varepsilon_1}{\bar{z} + \bar{z_1}} - \frac{\varepsilon_2}{\bar{z} + \bar{z_2}}.
\end{equation}

In the above equation,  $\varepsilon_1$ and $\varepsilon_2$ are the normalized lenses  masses,  with
 $\varepsilon_1 + \varepsilon_2 = 1$. The parameter $z$ is the \textbf{two-dimensional position written as the real and imaginary 
components of a complex number}. The $\omega$ is the relative position of the source at a specific time. \textbf{The bar over complex quantities indicates complex conjugation.}

\subsection{The semi-analytic method} 

Technically, to solve a lens equation with $n=2$, it is necessary to invert a 5th 
order polynomial and solve it to find the polynomial roots. To accomplish this task
we developed a model that uses a semi-analytic method to find  polynomial 
coefficients and  solutions \citep{wit90}. For the case where  the source is not close enough 
to the caustic-crossing region, we used the point source magnification method to solve 
and obtain  the light curve.

\section{Earth-mass like systems topology}
\label{section3}

Caustics modeling and microlensing  critical event curves depends fundamentally  on the apparent \textbf{semimajor} axis $s$ between the lenses, i.e., the host lens and the planet. Here we used  Einstein radius units $R_E$, and the mass fraction  as  \textbf{$q = m_2/m_1$, where  $m_2$  stands for the planet mass and $m_1$  mass of the star}.  The source's path is defined by 2 parameters, the impact parameter $u_0$ and the impact angle $\alpha$. The impact parameter $u_0$  represents the closest distance between the source and the host lens at the time $t_0$.   

In general,  binary systems caustics produce \textbf{close, resonant, and wide topologies} \citep{sch87,erd93}, and with limits varying  as a function of $s$ and $q$. For this case, the impact angle $\alpha$ is the angle between the source trajectory and the  $x$-axis of to the system. For the \textbf{binary-lens} case, the system lies in the $x$-axis. 

 For systems like our Sun-Earth system,  in terms of 
Earth-Sun mass ratio, we find $q = 3\times 10^{-6}$  and $s = 0.95969$, whereas the  \textbf{$m_1 = M_{\odot}$, $m_2 = M_\uplus$} and  $1 R_E = 1.0420$ AU. In such a system, a planet orbiting a \textbf{semimajor} axis of 1~AU \textbf{would lie} at the Einstein Ring limit.  Nevertheless, we can not ignore possible values of $s<$1~AU due to the fact that  for \textbf{this system} the  \textbf{semimajor} axis is the 
projected separation between the planet and its host star.  By considering systems with $q = 3\times 10^{-6}$ and  $s$ \textbf{as a free positive parameter}, two topologies \textbf{are more likely to} be obtained,  wide or close.  As presented by  \citep{erd93}, systems with such a wide topology satisfy the  condition 
\begin{equation}
s > \sqrt{\frac{(1+q^{\frac{1}{3}})^3}{1+q}}.
\label{widetopology}
\end{equation}

 For the interval  $0.1 < s < 0.95969$, our system can only be close. Thus,  to adjust the $u_0$
and $\alpha$ parameters in an efficient way,  we need to know the position   of the planetary caustic 
as a function of the  $s$ variation.

By analyzing equation \ref{widetopology}, we can conclude that a system with 
an Earth-Sun mass ratio can only be within a  wide topology if $s > 1.0217 R_E$.
On the other hand, as our system can only assume $0.1 R_E < s < 0.95969$, we can discard
the wide topology for systems like our own. Thus, to use microlensing path parametrization for Earth-like 
exoplanet detections  around solar mass stars, a deep analysis of the close topology case is necessary.

\begin{figure}
\centering
\includegraphics[width=\hsize]{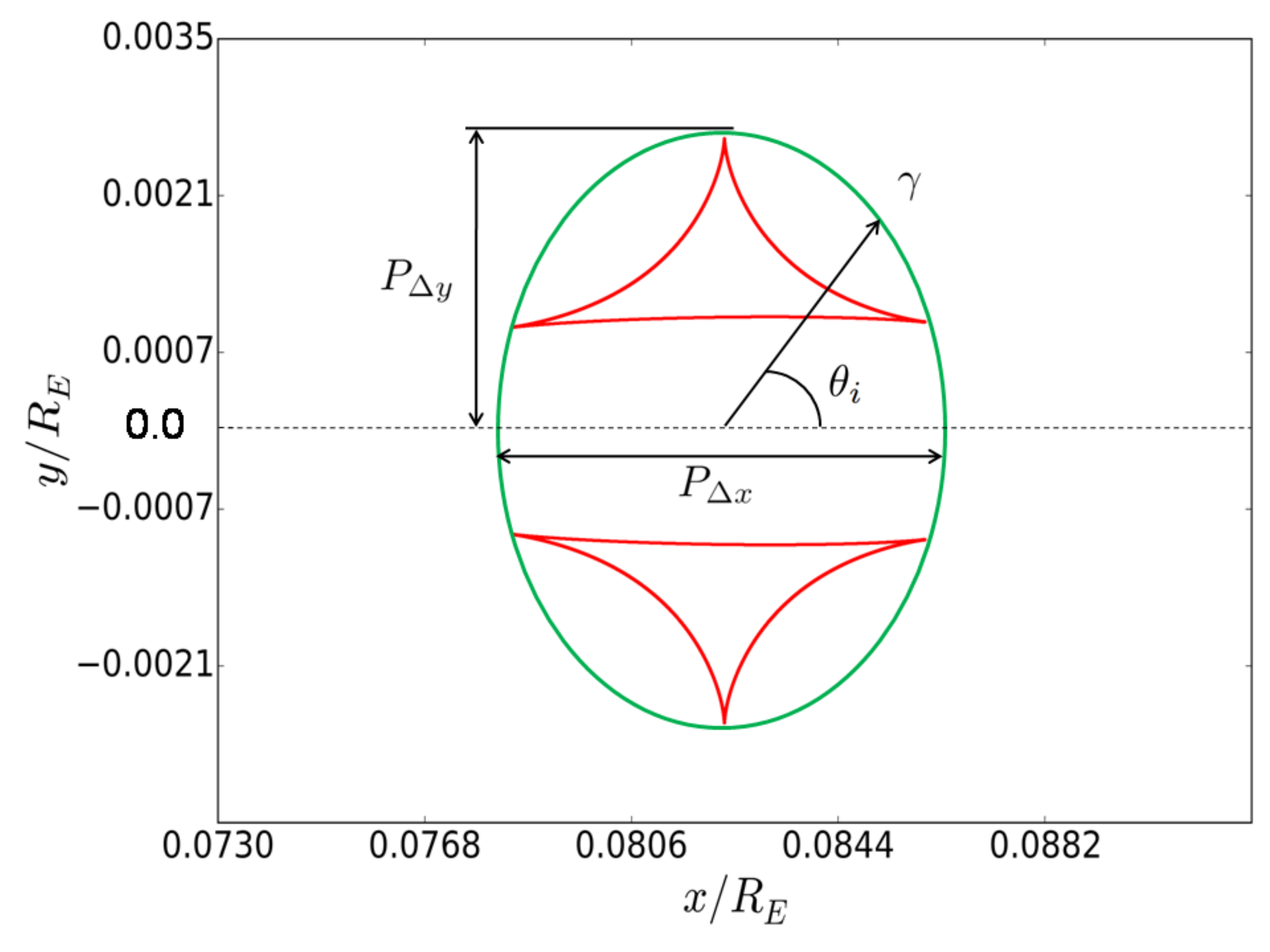}
\caption{Planetary caustic in detail with $q = 3.003467 \times {10}^{-6}$, $s = 
0.9597 E_R$, $u_0 = 0.0082 E_R$. The green ellipse is the influence area \textbf{defined} by
the equation \ref{AREA}}
\label{topology2}
\end{figure}

\subsection{close topology case}
\label{section3.1}

The close topology is formed by three caustics. A central caustic close to the primary lens
and two identical planetary caustics \textbf{on either side of the system axis} and  opposite side
of the planet. For a light curve of a source that passes close the central caustic 
and on the same side as the planet, we are able to detect only the main lens signature.
Following results by \citet{erd93}, we can define a such  close
topology system  when the condition below is satisfied

\begin{equation}
\frac{q}{(1+q)^2} < s^{-8} \left(\frac{1-s^4}{3}\right)^3.
\label{closecondition}
\end{equation}

In the above equation, for $q = 3\times 10^{-6}$, a system like our Sun-Earth system  can only be close if $s < 0.9893$. 
In order to set the region of influence, we need at this point,  to define  the planetary caustic characteristics for close systems.
Considering  $x$ as the  position of the planetary caustic, that  can be determined through the following 
equation \citep{han06}


\begin{equation}
X_{pc} = \frac{1}{1+q}\left(s - \frac{1-q}{s}\right),
\label{ondecausticplanet}
\end{equation}
where $X_{pc}$ is the the separation \textbf{between the primary} lens 
and the center of the planetary caustic. The equation \ref{ondecausticplanet} makes clear that
the smaller  $s$, the larger  the value of $X_{pc}$. By using this position $X_{pc}$
we were able to parametrized some geometrical proprieties of the system and also
to set the dependency of the source's path with the localization of the influence region around
the planetary caustic.  We can also link the position  $X_{pc}$  of the planetary caustic center  with
the impact parameter $u_0 $  by the following equation

\begin{equation}
u_0 = \frac{\left| {{s}^{2}}+q-1\right| \cdot \left| \mathrm{tan}\left(
\mathit{\ensuremath{\alpha}}\right) \right| }{\left| q+1\right| \cdot \left| s\right|
\cdot \sqrt{{{\mathrm{tan}\left( \mathit{\ensuremath{\alpha}}\right) }^{2}}+1}}.
\label{muvalue}
\end{equation}

\begin{figure}
\includegraphics[width=\hsize]{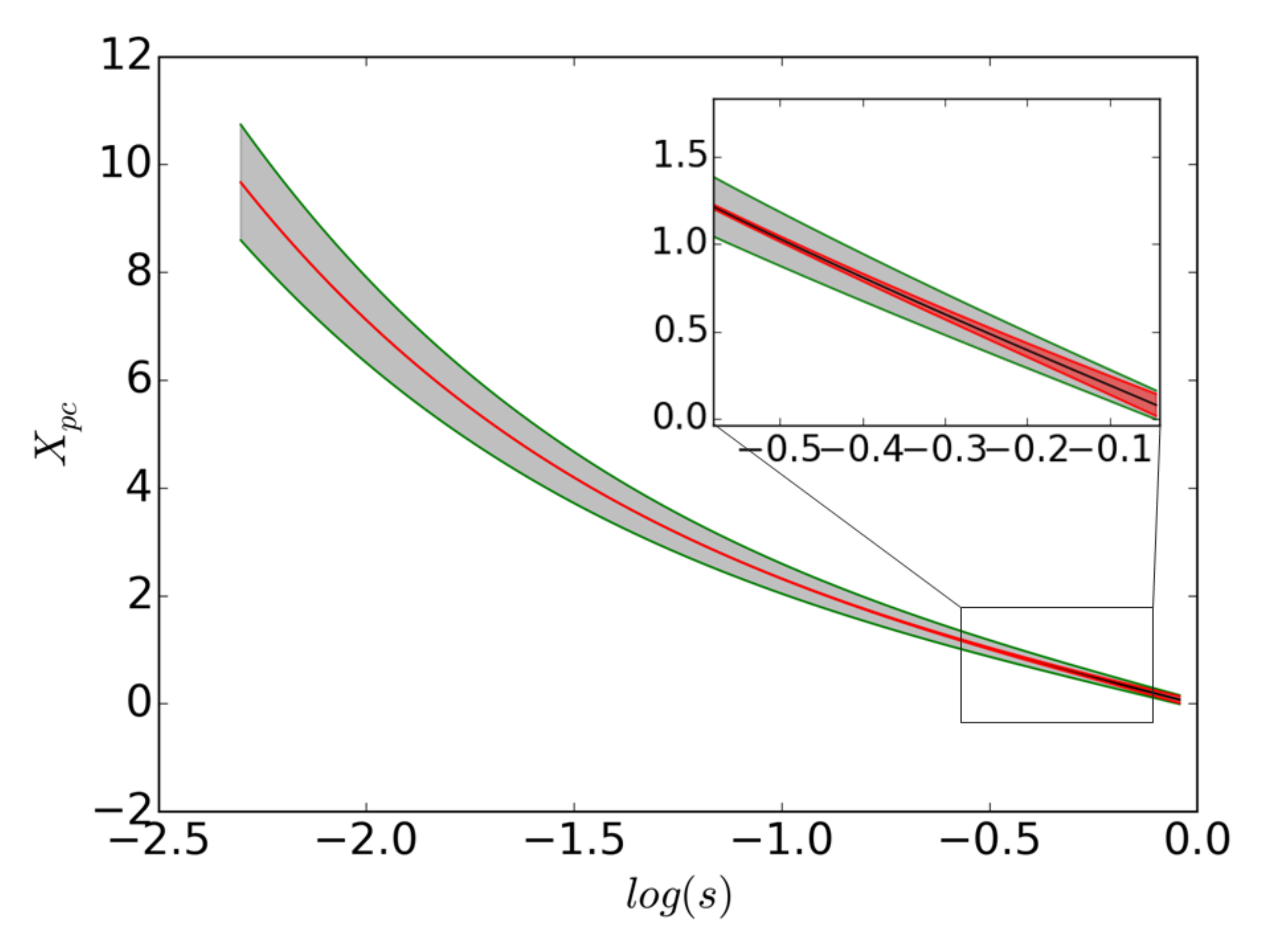}
\caption{$X_{pc}$ as a function of   $\log{s}$ for our \textbf{adopted} system with
$q = 3.003467\times 10^{-6}$ and $0.95969 > s > 0.1$. The gray region is $2P_{\Delta y}$
and the red region is $P_{\Delta x}$ (both multiplied by twenty for better visualization)}
\label{Area}
\end{figure}

To better describe the entire region of interest   we need to geometrically describe the entire area containing the planetary caustic.  For that,  following the geometry of the
 problem, we found values for $P_{\Delta x}$ and $P_{\Delta y}$, (Figure \ref{topology2}) written below

\begin{equation}
P_{\Delta x} = \frac{3}{2} s^{3} \sqrt{3}\sqrt{q},
\label{pdx}
\end{equation}
\begin{equation}
P_{\Delta y} = 2 \frac{\sqrt{q}}{s \sqrt{s^{2}+1}}.
\label{pdy}
\end{equation}

For close topologies in this regime, the planetary caustic can be enclosed by
an ellipse independent of its size. Thus the size of the influence area,
which contains the planetary caustic, can be defined through an ellipse area
$ \pi a b$, with $a = P_{\Delta x}/2$ 
and $b = P_{\Delta y}$. Thus, the influence area that define the region containing  the planetary caustic is 
\begin{equation}
A = \pi \frac{P_{\Delta x}}{2} P_{\Delta y}.
\label{area1}
\end{equation}

Entering the equations \ref{pdx} and \ref{pdy} into the equation \ref{area1}, we determined the  \textbf{area}  $A$ that contains the planetary 
caustic as presented by the green ellipse in the figure \ref{topology2}, and now as a function of  $q$ and $s$

\begin{equation}
A = \frac{\gamma^2 \pi s^2 \sqrt{3} q}{\sqrt{s^2 + 1}},
\label{AREA}
\end{equation}
where $\gamma$ is a scalar factor for the size of the area  which contains the planetary caustic.
For the particular case of   $\gamma$ equal to $1$, figure \ref{topology2},  such area fits perfectly the planetary caustic.\\

By analyzing  figure \ref{Area} we find that, for systems with close topology,
the distance $X_{pc}$ increases as $s$ decreases. We can also see, based on equations
\ref{pdx} and \ref{pdy},  that  $P_{\Delta x}$ drastically decreases 
 and  $P_{\Delta y}$ increase when $s$ approaches the origin. Equation \ref{AREA} leads to the conclusion
that the area of the planetary caustic overall decreases when $s$ approaches to  origin.
Thus, even with $X_{pc}$ getting bigger when $s$ decreases, the  total area is not enough 
 for any possible detection. Figure \ref{Area}  leads to the conclusion that $2 P_{\Delta y}$ and $P_{\Delta x}$ approaches to same value when
$s$ approaches to $1$.\\


To link the source path with the \textit{Caustic Region Of INfluence} (CROIN) as described by \cite{penny14}, we define all the points on the ellipse using the equations \ref{pdx} and \ref{pdy} as

\begin{equation}
X_{ip} = \gamma P_{\Delta x} cos(\theta_i) + X_{pc},
\label{xip}
\end{equation}
\begin{equation}
Y_{ip} = \gamma P_{\Delta y} sin(\theta_i).
\label{yip}
\end{equation}

If we evolve $\theta_i$ from $0$ to $2\pi$ in the equations above, we define the perimeter of the 
ellipse of area $A$,  for the close topology case. Now,  we can define the   parameterization of the source path to the close topology case  by the next  equation

\begin{equation}
u_i = \frac{\left|tan(\alpha)X_{ip} - Y_{ip} \right|}{\sqrt{{{\mathrm{tan}
\left( \mathit{\ensuremath{\alpha}}\right) }^{2}}+1}}.
\label{mui}
\end{equation}

By setting $\gamma = 1$, and varying $\alpha$ from $0$ to $2 \pi$, 
we \textbf{obtain all} values of $u_0$  from the equation \ref{mui} with  the path of the source
always  passing by the planetary caustic vicinity. Thus, to explore all the possible light curves for our Earth-Sun model, we need to
vary  $\gamma$, $\alpha$ and $\theta_i$.  Furthermore we know from \citep{1986ApJ...304....1P} that, when analysing a microlensing event, the parameters $t_E$, $t_0$ and $u_0$ are
the firsts to be \textbf{established} from the single-lens model. It is more interesting here to  parametrize $\alpha$ in respect to $u_0$, because the 
impact parameter $u_0$ is
already set to a small error from the single-lens model. We can note that the equation \ref{mui} is impossible to be inverted in terms of $\alpha(u_0)$, so we 
need to find another method to express the parametrization of $\alpha$ in respect to
$u_0$.

\begin{figure}
\includegraphics[width=\hsize]{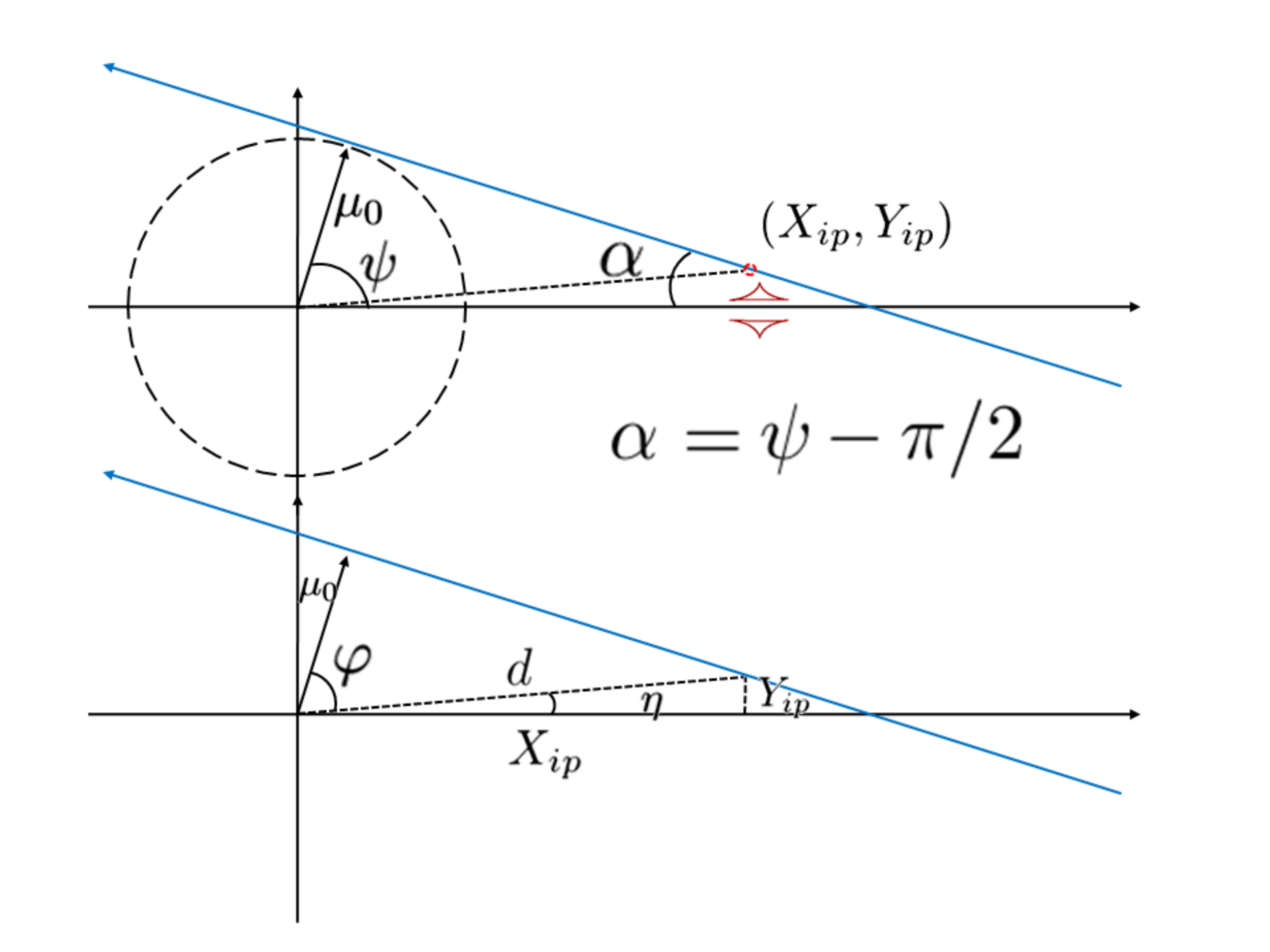}
\caption{Top panel: Topology of a close system showing the point of interest from equations 18 and 19. Bottom panel:  the geometry of the system with relative angles.}
\label{tangent}
\end{figure}

To achieve that we need to find a function $\alpha(u_0)$ which depends only on the position
of interest given by $X_{ip}$ and $Y_{ip}$, and the impact parameter $u_0$. 
By analysing the geometry (figure \ref{tangent}), we get  $d = (X_{ip}^2+Y_{ip}^2)^{1/2}$,
$\eta = acos(X_{ip}/d)$ and $\varphi = acos(u_0/d)$. The impact angle $\alpha$ is $\psi - \pi/2$
with $\psi$ being the sum of $\eta$ and $\varphi$. Then, we can simplify our new function $\alpha$
as:

\begin{equation}
\alpha(u_0) = acos\left(\frac{u_0}{\sqrt{X_{ip}^2+Y_{ip}^2}}\right) -
asin\left(\frac{X_{ip}}{\sqrt{X_{ip}^2+Y_{ip}^2}}\right).
\label{alphapath}
\end{equation}

Notice that equation \ref{alphapath} depends solely on $q$, $s$ and $u_0$ and can 
fill the area of the planetary caustic by varying $\gamma$.
This parametrization \textbf{only covers} the set of $q$ and $s$ that generate close topologies.
For values out of this range (wide or resonant), we can not use this parametrization.
The evolution of the impact angle $\alpha$ was computed when different initial $u_0$ 
was set as  $s = 0.95969$, $0.85$, $0.75$, $0.6$ and $0.4$. For all cases the impact 
parameter $u_0$ must be smaller than the position of the planetary caustic or else the path of
the source will not pass through the region of influence.\\

\begin{figure}
\includegraphics[width=\hsize]{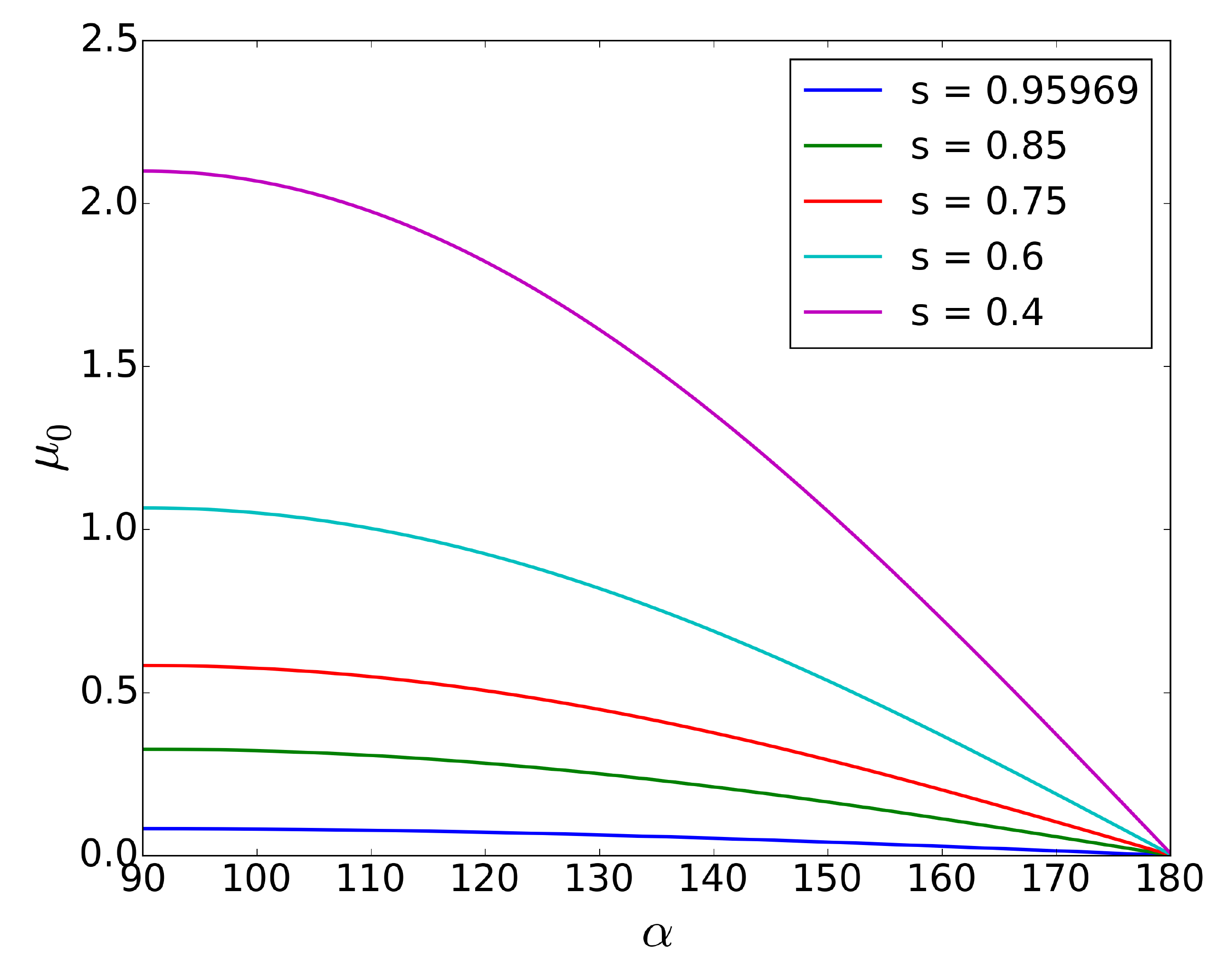}
\caption{Evolution of the impact angle $\alpha$ when different initial $mu_0$ is set for
 $s = 0.95969, 0.85, 0.75, 0.6$ e $0.4$.}
\label{mu0xalpha}
\end{figure}

Figure \ref{mu0xalpha} presents the evolution of the impact angle $\alpha$ when different initial $u_0$ 
is set to  $s = 0.95969$, $0.85$, $0.75$, $0.6$ and $0.4$. We can see in all cases that the impact 
parameter $u_0$ must be smaller than the position of the planetary caustic or else the path of
the source will not pass through the region of influence.\\

From the figure \ref{mu0xalpha} and relative equation \ref{muvalue} we see that, as
$\alpha$ approaches $90^\circ$ (perpendicular with the lens axis) the value of $u_0$ increases.
That happens because, in order to the source's path to cross the interest region in 
$X_{pc}$, $u_0$ \textbf{needs to} be $0$ so that $\alpha = 2 \pi$ and if the path is perpendicular,
with $\alpha = \pi /2$, than $u_0$ must be set to the value of $X_{pc}$.
According to \cite{penny14}, this kind of parametrization can greatly
accelerate the simulation of light curves in the search for low-mass planets,
but at the cost of passing by possible detections in unlikely topologies.

\subsection{Light curves for close  systems}\label{section3.2} 
\begin{figure}[h!]
\vspace{0.8cm}
\includegraphics[width=\hsize]{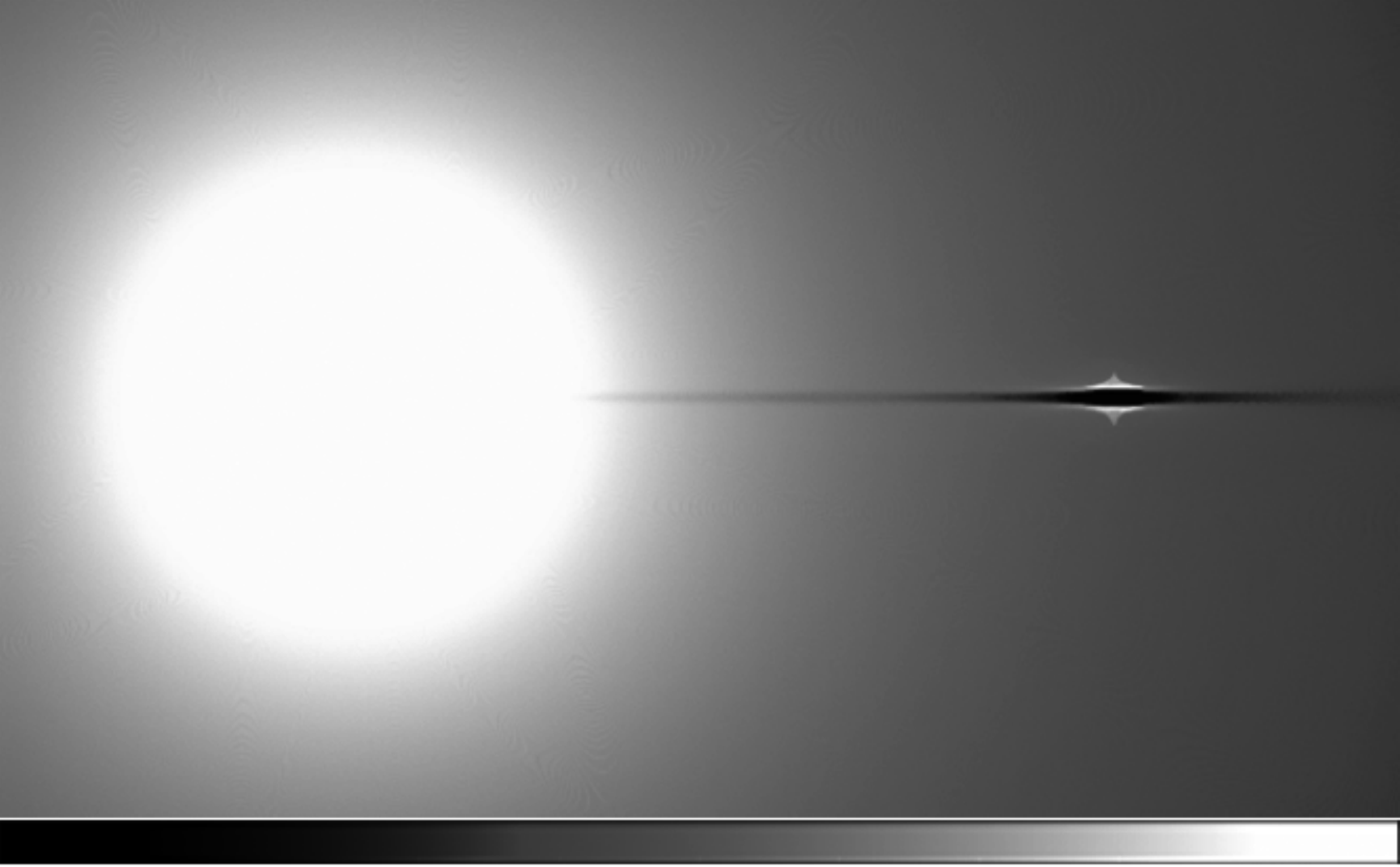}
\caption{Magnification map of a system with $q = 3.003467 \times {10}^{-6}$ and 
$s = 0.9597 E_R$. The color bar shows
arbitrary values from low magnification(black) to high magnification (white).}
\label{magmap22}
\end{figure}

\begin{figure}
\includegraphics[width=\hsize]{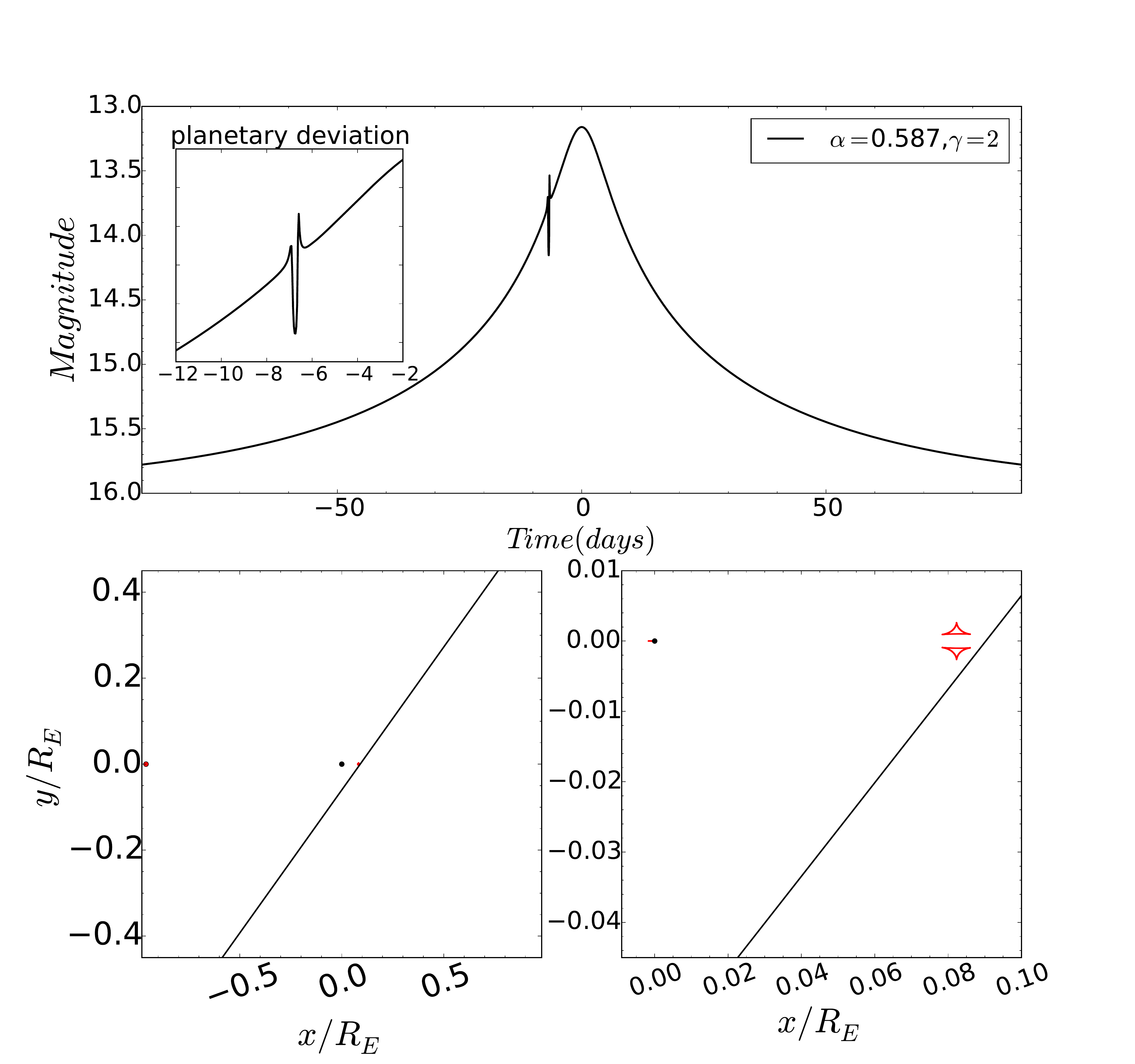}
\caption{Top panel: light curve model for Sun-Earth system with $q = 3.003467\times 10^{-6}$ and $s = 0.95969$
with a close-up at the planetary deviation. Bottom panel: left panel shows a wide view of the system with the path in
blue; right panel is a close-up of the planetary caustic.}
\label{lightcurve1}
\end{figure}

Once we have the parametrization of $\alpha(u_0)$ and $u_0(\alpha)$ in respect to
the positions $X_{ip}$ and $Y_{ip}$, we can generate all the light curves within the 
region of interest by varying $\gamma$ in the equations \ref{xip} e \ref{yip}.  Figure \ref{lightcurve1} shows a light curve of a system that mimics our own Sun-Earth system with
$q = 3.003467\times 10^{-6}$, $s = 0.95969$ and path parameters as $u_0 = 0.15$, and $\alpha = 0.587$.  
We note a negative magnification at the planetary crossing region due to the source passing between the two planetary caustics. 
This negative magnification can be better visualized in 
the figure \ref{magmap22}.\\

Figure \ref{magmap22} shows the magnification map of  our system created by using a 
$5000\times5000~pixels$ grid with arbitrary values for magnification. We can see
that the central lens is responsible for almost all magnification of the source. 
The deviation due to the planetary caustic is  negative between the two caustics but
\textbf{also presents} a  positive magnification at the crossing caustic regions.\\

\begin{figure*}
\includegraphics[width=\textwidth]{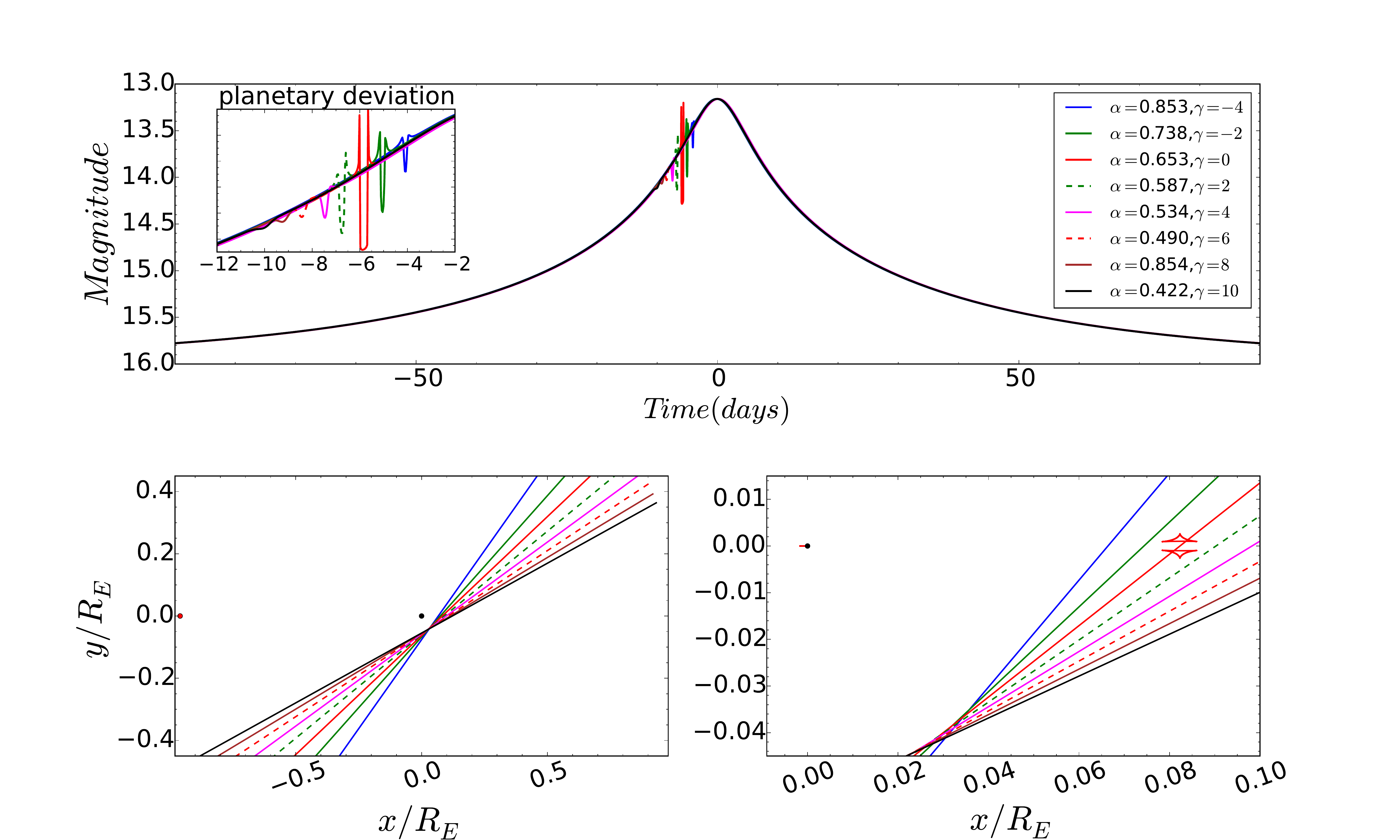}
\caption{Top panel: 8 superposed light curves with $-4 < \gamma 10$ for our 
simulated system; the planetary deviation panel shows a close-up at all the
planetary signals. Bottom panels: the left panel shows the different source path
for each $\gamma$ on top of the topology of the system; the left panel shows
the planetary caustic region in close-up.}
label{lightcurve3}
\end{figure*}
By using the equation \ref{alphapath} we generated several light curves by setting a fixed value for $u_0 = 0.15$ and varying $\gamma$ from $-4$ to $10$ thus, evolving the 
values of $\alpha$ from $0.422$ (black line) to $0.853$ (blue line). From our parametrization, we can see in the figure \ref{lightcurve1} \textbf{that the} overall aspect of the light curve for a  single-lens case is preserved and that the end of all possible planetary deviations  are 
superposing the same line. Thus, given the initial parameters $t_0$, $t_E$ and $u_0$ from a single-lens approximation scenario
we are able to generate all possible light curves that could present a detectable planetary deviation. The detection
itself depends on the observational cadence.\\

To demonstrate the computational efficiency and an increase of the precision from our method we performed a computational experiment and produced synthetic systems with the following parameters:

\begin{itemize}
\item {\it cadence}:  24 daily photometric measurements
\item {\it tobserv}: observational duration: 90 days
\item {\it number of points}:  cadence*tobserv
\item {\it u}: impact parameter  = 0.05  
\item {\it alpha}: inclination of impact  = -2.489 
\item {\it q}: mass fraction  = 3.003467e-6 
\item {\it s}: \textbf{normalized projected separation} = -0.95969  
\item {\it tE}:time in days to cross the Einstein radius  = tobserv / 2 
\end{itemize}


The experiment \textbf{generates}  synthetic data with 4320 photometric points spread along 90 days and with a record every 1 hour. We apply a Gaussian noise error of $0.5\%$ in the photometric measurement.

\begin{figure}
\includegraphics[width=\hsize]{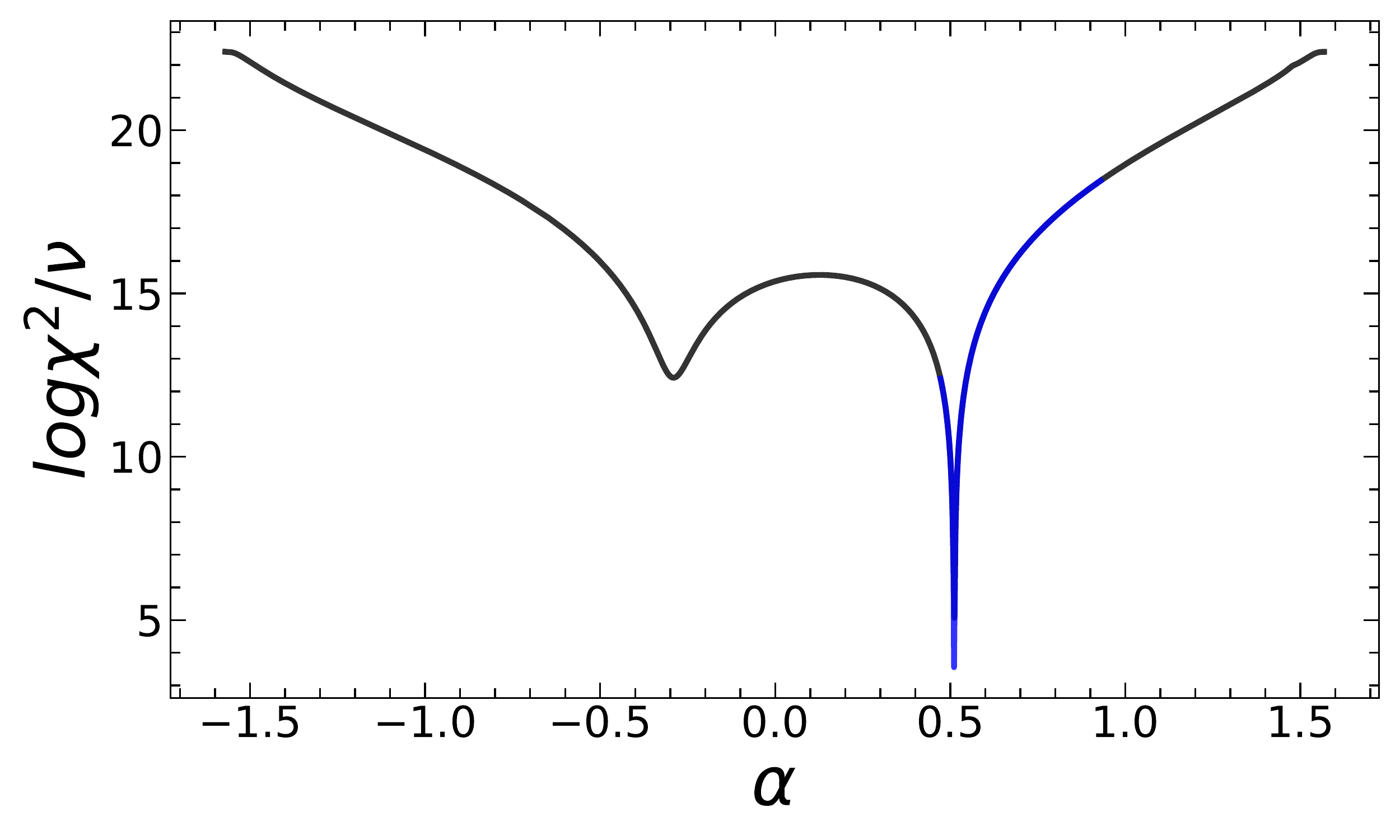}
\caption{$\chi^2$ diagram showing the reduced impact angle  $\alpha$. The blue line represents the search process by using our parametrization.}
\label{figura1}
\end{figure}

Based on the synthetic light curves,  a systematic search for parameters was performed by setting $q$ and $s$ as our simulated system. Then, the same  systematic search for parameters was performed by using our new parameterization.  On the conventional method, we need to cover the dispersion of the $q$ and $s$ parameters, and we need also to cover the impact  angle variation  as $2 \pi > \alpha > 0$. We set all other parameters as described above  and  a search now only on the $\alpha$. The denser variation of $\alpha$ \textbf{gives more} accuracy to the result. We run the code to cover $2\pi > \alpha > 0$ with 1000 points.
After that, we run the code using our model, varying the parameter $\gamma$ from  $-5$ to $8$, with the same points quantity. We show in the figure \ref{figura1} the comparative performance result between our  method and the conventional one.
From the figure \ref{figura1} the blue line represents the search process by  using our model.  We  can see that,  
the search performed with the conventional model (black line) \textbf{covers} some unnecessary regions of 
the \textbf{alpha domain}. A second aspect is that in addition to \textbf{region covered}, we have less resolution in the regions of smaller $\chi^2$. We see that  $\alpha$ and $u_0$  parameterization with respect to CROIN, forces the search to be focused  only in the region that it would be possible to detect our kind of interested system.\\

As far as the parameterization deals with a focused search process,  
its efficiency is mainly controlled by the ratio between the global 
search $\alpha$ (conventional model) and the focused one  $\gamma$. 
Based on that, we conclude that, for this particular case, we arrive at the same result 
with a precision rate    $5.2 $ higher, and  by using 1000 points in a range of 0.6 rad, instead $\pi$ on the conventional one. For the same density of points and accuracy, \textbf{our methods is 5 times faster to converge to the best fit and this is one of advantages}.

\section{SUMMARY AND DISCUSSION}
\label{summary}


We analyzed a set of simulations constructed to search Earth-like exoplanet around solar mass stars.
Our simulations involved a parameter search on Sun-Earth models created using the semi-analytical 
method. We find that all solutions involving close topologies are not 
degenerated and since we are searching only around the region of interest.
Our parametrization efficiency is mainly controlled by the ratio between the global search $\alpha$  and  $\gamma$. 
Based on that, we arrive for our simulated case,  the  same correct result  with a precision rate $5.2 $ higher.
For the same density of points and accuracy, our methods is 5 times faster.
 For a system with mass fraction and \textbf{semimajor} axis apparent similar to our
Sun-Earth system and $t_E = 90$ days, we find that the planetary deviation takes  about 1 day and can be observed by \textbf{
a high cadence surveys.  The majority of microlensing events has typical timescale of about 20 days. LSST with first light planed to 2019,  does not plan to survey the bulge, but in any
case,  has  cadence enough of about $1/4~days$ for 
field events and could in principle trigger follow-up observations to 
search for planets. WFIRST planed to be launched in 2024 has
appropriated cadence and will observe  the bulge and other field}.

We find that Sun-Earth analog observed system  will present a close topology (for \textbf{semimajor} axis close to $1~AU$) with doubled identical caustics on the other side of the planet. We also concluded that the ellipse around the planetary caustic decreases exponentially as $s$ increases.
We find that if the \textbf{semimajor} axis is equal to $1~AU$, then the deviation of the light curve
from the single-lens case will last for about one day (for $t_E = 90$ days). The new
values for $X_{ip}$ and $Y_{ip}$ are implemented within the new parametrization of $\alpha(u_0)$
and can easily be integrated in the parameters search with $\gamma$ dictating the evolution of
$\alpha$ once we have define a fixed $u_0$.\\ 

\textbf{We would like to thank the anonymous referee, whose important suggestions, greatly improved the paper without a doubt. We thanks to CAPES and the Federal University of Rio Grande do Norte (UFRN) for financial support. JDNJr acknowledges financial support by Brazilian CNPq PQ~1D grant n$^{\circ}$310078/2015-6.}


\begin{thebibliography}{}

\bibitem[Albrow et al. (2001)]{alb01} Albrow, M. D., et al. 2001, Limits on the Abundance of Galactic Planets From 5 Years of PLANET Observations, Astrophys. J. Lett., 556, L113

\bibitem[Alcock et al.(1995)]{1995PhRvL..74.2867A} Alcock, C., Allsman, R.~A., Axelrod, T.~S., et al.\ 1995, Physical Review Letters, 74, 2867

\bibitem[Beaulieu et al.(2011)]{2011IAUS..276..349B} Beaulieu, J.-P., Bennett, D.~P., Kerins, E., \& Penny, M.\ 2011, The Astrophysics of Planetary Systems: Formation, Structure, and Dynamical Evolution, 276, 349 

\bibitem[Bennett \& Rhie(1996)]{1996ApJ...472..660B} Bennett, D.~P., \& Rhie, S.~H. 1996, apj, 472, 660

\bibitem[Bozza(2000)]{2000A&A...359....1B} Bozza, V. 2000, aap, 359, 1

\bibitem[Cassan et al.(2012)]{2012Natur.481..167C} Cassan, A., Kubas, D., Beaulieu, J.-P., et al.\ 2012, \nat, 481, 167

\bibitem[do Nascimento et al.(2016)]{2016ApJ...820L..15D} do Nascimento, J.-D., Jr., Vidotto, A.~A., Petit, P., et al.\ 2016, \apjl, 820, L15 

\bibitem[Einstein(1936)]{1936Sci....84..506E} Einstein, A.\ 1936, Science, 84, 506 

\bibitem[Erdl \& Schneider(1993)]{erd93} Erdl, H., \& Schneider, P. 1993, aap, 268, 453

\bibitem[Gaudi et al. (2002)]{gau02} Gaudi, B. S., et al. 2002, Microlensing Constraints on the Frequency of Jupiter-Mass Companions: Analysis of 5 Years of PLANET Photometry, Astrophys. J., 566, 463

\bibitem[Gould \& Loeb(1992)]{1992ApJ...396..104G} Gould, A., \& Loeb, A.\ 1992, apj, 396, 104

\bibitem[Gould et al.(2014)]{2014Sci...345...46G} Gould, A., Udalski, A., Shin, I.-G., et al.\ 2014, Science, 345, 46

\bibitem[Griest et al.(2011)]{2011PhRvL.107w1101G} Griest, K., Lehner, M.~J., Cieplak, A.~M., \& Jain, B.\ 2011, Physical Review Letters, 107, 231101 

\bibitem[Han \& Gould,(1995)]{han95} Han, C., Gould, A. 1995, ApJ, 447, 53

\bibitem[Han (2006)]{han06}Han, C. 2006, ApJ, 638, 1080

\bibitem[Liebes(1964)]{1964PhRv..133..835L} Liebes, S.\ 1964, Physical Review, 133, 835

\bibitem[Mao \& Paczynski(1991)]{1991ApJ...374L..37M} Mao, S., \& Paczynski, B.\ 1991, apjl, 374, L37

\bibitem[Mayor \& Queloz(1995)]{1995Natur.378..355M} Mayor, M., \& Queloz, D.\ 1995, \nat, 378, 355

\bibitem[Paczynski(1986)]{1986ApJ...304....1P} Paczynski, B.\ 1986, \apj, 304, 1

\bibitem[Penny(2014)]{penny14} Penny, M.~T. 2014, apj, 790, 142 

\bibitem[Rhie(1999)]{1999astro.ph..9433R} Rhie, S.~H.\ 1999, arXiv:astro-ph/9909433 

\bibitem[Rhie et al. (2000)]{rhi00}    Rhie, S. H. et al. 2000, On Planetary Companions to the MACHO 98-BLG-35 Microlens Star, Astrophys. J., 533, 378

\bibitem[Schneider \& Weiss(1987)]{sch87} Schneider, P., \& Weiss, A. 1987, aap, 171, 49

\bibitem[Shvartzvald et al.(2017)]{2017ApJ...840L...3S} Shvartzvald, Y., Yee, J.C., Calchi Novati, S., et al. 2017, apjl, 840

\bibitem[Sumi et al.(2003)]{2003ApJ...591..204S} Sumi, T., Abe, F., Bond, I.~A., et al.\ 2003, \apj, 591, 204

\bibitem[Sumi et al (2006)]{sum06} Sumi, T., et al. 2006, Microlensing Optical Depth toward the Galactic Bulge Using Bright Sources from OGLE-II, Astrophys. J., 636, 240

\bibitem[Udalski et al.(2018)]{2018AcA....68....1U} Udalski, A., Ryu, Y.-H., Sajadian, S., et al.\ 2018, \actaa, 68, 1

\bibitem[Witt H. J (1990)]{wit90} Witt H. J., 1990, Astronomy and Astrophysics (ISSN 0004-6361)

\bibitem[Witt \& Mao (1995)]{wit95}Witt, J., Mao, S., 1995, ApJ, 447, 105

\bibitem[Yee et al.(2009)]{2009ApJ...703.2082Y} Yee, J.C., Udalski, A., Sumi, T., et al. 2009, apj, 703, 2082

















\end{thebibliography}
\end{document}